# Data class-specific all-optical transformations and encryption

*Bijie Bai [1,2,§], Heming Wei[3,§], Xilin Yang[1,2], Deniz Mengu[1,2], and Aydogan Ozcan[1,2,4\*]*

[1]Electrical and Computer Engineering Department, University of California, Los Angeles, CA, 90095, USA.

[2]Bioengineering Department, University of California, Los Angeles, 90095, USA.

[3]Key Laboratory of Specialty Fiber Optics and Optical Access Networks, Joint International Research Laboratory of Specialty Fiber Optics and Advanced Communication, Shanghai University, Shanghai 200444, China

[4]California NanoSystems Institute (CNSI), University of California, Los Angeles, CA, USA.

[§] Equal contribution

[\*] Correspondence: Aydogan Ozcan. Email: ozcan@ucla.edu

**Abstract**

Diffractive optical networks provide rich opportunities for visual computing tasks since the spatial information of a scene can be directly accessed by a diffractive processor without requiring any digital pre-processing steps. Here we present data class-specific transformations all-optically performed between the input and output fields-of-view (FOVs) of a diffractive network. The visual information of the objects is encoded into the amplitude ($A$), phase ($P$), or intensity ($I$) of the optical field at the input, which is all-optically processed by a data class-specific diffractive network. At the output, an image sensor-array directly measures the transformed patterns, all-optically encrypted using the transformation matrices pre-assigned to different data classes, i.e., a separate matrix for each data class. The original input images can be recovered by applying the correct decryption key (the inverse transformation) corresponding to the matching data class, while applying any other key will lead to loss of information. The class-specificity of these all-optical diffractive transformations creates opportunities where different keys can be distributed to different users; each user can only decode the acquired images of only one data class, serving multiple users in an all-optically encrypted manner. We numerically demonstrated all-optical class-specific transformations covering $A \to A$, $I \to I$, and $P \to I$ transformations using various image datasets. We also experimentally validated the feasibility of this framework by fabricating a class-specific $I \to I$ transformation diffractive network using two-photon polymerization and successfully tested it at 1550 nm wavelength. Data class-specific all-optical transformations provide a fast and energy-efficient method for image and data encryption, enhancing data security and privacy.

# 1. Introduction

In recent years, optical computing has re-gained attention due to its potential advantages in speed, energy efficiency, and scalability. Various photonic devices have been developed to implement linear transformations using, e.g., Mach–Zehnder interferometers[1], optical tensor cores[2], complex optical media[3], and diffractive deep neural networks[4,5] ($D^2NNs$). The latter, $D^2NNs$ or diffractive optical networks[4,6], form an emerging platform that uses deep learning methods to design diffractive surfaces for all-optical or optoelectronic (hybrid) computing. Once they are fabricated (after their optimization), diffractive networks can perform complex computer vision tasks, such as image classification[4,7–9], hologram reconstruction[10], quantitative phase imaging (QPI)[11], imaging through diffusers[12], and others[13–22] using light diffraction through passive structured layers, one layer following another within a compact volume. Especially for visual computing tasks, diffractive networks provide unique advantages compared to other photonic computing hardware; diffractive networks do not need to pre-process visual information since they have direct access to and can process the 2D and 3D spatial information of a scene, encoded by e.g., amplitude, phase, polarization and spectrum of the input light. One example of this is optical phase recovery, which normally uses interferometric or holographic imaging hardware for encoding the phase information of an object into an intensity pattern, which is then digitized, stored/transmitted, and reconstructed using, e.g., an electronic neural network[23] or an algorithm to retrieve the encoded phase information of the object. Diffractive networks, on the other hand, can directly process the phase information of a sample without any hologram acquisition step or digital phase recovery algorithms, and recover the quantitative phase images of specimens using diffraction of light through trained transmissive layers, performing all-optical phase recovery and holographic image reconstruction[11]. These diffractive visual processors do not consume power (except for the illumination light) and are typically very compact, spanning axially, e.g., ~50-300×λ, where λ is the illumination wavelength. This thin architecture also makes them ultra-fast, completing their inference or image processing/reconstruction tasks as the light is transmitted through a passive optical volume that is structured, layer by layer, at the wavelength scale with a lateral feature size of ~λ/2. Despite these advantages, 3D fabrication and free-space alignment of the resulting diffractive layers of a deep $D^2NN$ architecture, especially in the visible and IR bands, are generally challenging[18,20] due to the tight tolerance requirements in shorter wavelengths. Furthermore, reconfiguring diffractive layers for different visual computing tasks requires the fabrication of new diffractive layers or the use of relatively bulky and slow spatial light modulators (SLMs) as part of its set-up, which presents challenges for applications where the input data stream or the desired task of interest change over time.

Here we expand on the family of visual transformations enabled by diffractive networks and present D²NN-based data class-specific image transformations that are all-optically performed between the input and output fields-of-view (FOVs) of a diffractive network. At the input FOV, we consider the diffraction-limited optical information to be encoded in one of the following schemes: amplitude-only ($A$), phase-only ($P$), and intensity-only ($I$). Based on these three schemes, we analyze all-optical transformations that perform data class-specific operations between the input and output FOVs covering (1) $A \rightarrow A$, (2) $I \rightarrow I$, and (3) $P \rightarrow I$ transformations (see Table 1). These transformations are selected so that an image sensor-array positioned at the output FOV can directly detect the all-optical transformation results, revealing the output amplitude or intensity patterns. In each one of these three different schemes considered here, there are real-valued non-negative and normalized transformation matrices ($T_m$) that are individually assigned as a target transformation between the input and output FOV for each data class of interest, where $m = 1, 2, ..., N$ (see Fig. 1a), and $N$ refers to the total number of data classes of interest. In general, these data-class-specific transformations are randomly selected and $T_p \neq T_k$ unless $p = k$.

For example, consider the family of $P \rightarrow I$ transformations between the input and output FOVs (see Table 1 and Fig. 1b); this constitutes a transformation between the 2D phase information of an input phase-only object ($i_n = e^{j2\pi\phi_n}$, with $\phi_n \in [0, 1]$) and the output intensity distribution ($I$), where $n$ refers to the data class that this input object belongs to. To approximate these $P \rightarrow I$ transformations for each data class, the linear optical forward model of the diffractive network is trained using deep learning to accordingly process the phase-only input fields, i.e., at the output FOV, we have $D^2NN\{e^{j2\pi\phi_n}\} = \mathbf{D}i_n$. For different data classes, $n = 1, 2, ..., N$, the diffractive network is trained to approximate the desired set of arbitrary transformations, $T_m$, such that if $n = m$, we have at the output FOV, $|\mathbf{D}i_n|^2 \propto T_m\phi_m$, performing the desired $P \rightarrow I$ transformation using $T_m$ assigned to data class $m$. We have similar goals for the $A \rightarrow A$ and $I \rightarrow I$ transformations, i.e., if and only if $n = m$, $|\mathbf{D}i_n| \propto T_m|A_m|$ and $|\mathbf{D}i_n|^2 \propto T_m|A_m|^2$, respectively.

We demonstrated these different class-specific optical transformations, $A \rightarrow A$, $I \rightarrow I$, and $P \rightarrow I$, with separately trained D²NNs using the images from the EMNIST dataset[24], the QuickDraw dataset[25], and the Fashion MNIST dataset[26]. These class-specific optical transformation schemes with different encoding methods ($A, I,$ or $P$) were numerically investigated to show that our class-specific transformation framework can be flexibly utilized in different settings with different types of objects. Our results demonstrate that class-specific D²NNs can perform the desired transformations assigned to different input data classes, resulting in all-optically encrypted measurements at the output image sensor-array. Only with the correct cryptographic keys, i.e., the class-specific inverse transformations, $T_m^{-1}$, the D²NN-encrypted images can be restored to reveal the original image information. Applying the other inverse transformations to the non-matching data classes ($n \neq m$) will lead to uninterpretable images (Fig. 2), confirming the success of class-specific all-optical transformation D²NNs.

In addition to our numerical analyses, we also experimentally demonstrated the proof of concept of this class-specific all-optical image transformation scheme by building an $I \rightarrow I$ transformation diffractive optical network that operates at λ=1550 nm. This class-specific D²NN was trained using the MNIST dataset for specifically imaging handwritten digits "2" at its output FOV and all-optically erasing the other data classes such that for $m = n = 3$ (the handwritten digit "2" class), we have $T_m \approx \bar{I}$, where $\bar{I}$ is the identity matrix, and for all the other data classes ($n \neq 3$), the object information cannot be retrieved at the output FOV. Using two-photon polymerization-based 3D printing[27], we fabricated and tested a 2-layer D²NN to experimentally demonstrate the success of this class-specific diffractive network under 1550 nm illumination.

The presented data class-specific transformation network operates entirely based on light propagation through passive transmissive layers, which does not require any external computing power except for the illumination light. Since the class-specific transformation is achieved all-optically through the modulation of the objects' wave field, our framework is especially attractive for image and data encryption, enhancing data security and privacy. It permits encrypted imaging of desired classes of input objects, with a set of transformation matrices that are data-class specific, and for all the other types of undesired objects, it all-optically erases the information of the input. The class-specificity of these transformations also opens up opportunities where different keys ($T_m^{-1}$) can be distributed to different users, and each user can only decode the acquired images of only one data class (see e.g., Fig. 2), helping the system serve multiple users in an all-optically encrypted manner. All-optical class-specific transformations through D²NNs will pave a viable avenue to further the development of image and data security solutions and image processing devices that are fast, task-specific, and energy-efficient.

## 2. Results

### 2.1. Data class-specific all-optical transformations

As summarized in Fig. 1b and Table 1, the presented all-optical class-specific transformation network can be designed to operate under different working schemes: **(1)** amplitude to amplitude transformation ($A \rightarrow A$); **(2)** intensity to intensity transformation ($I \rightarrow I$); and **(3)** phase to intensity transformation ($P \rightarrow I$). In all of these selected families of transformations, an optoelectronic sensor-array, e.g., a complementary metal–oxide–semiconductor (CMOS) image sensor or its equivalent, would be sufficient to directly record the transformation result (intensity or amplitude information) at the output FOV without any additional hardware or computing. Given a real-valued normalized image ($G_n \in [0, 1]$) that belongs to the data class $n$, the image is first encoded into the phase, amplitude, or intensity channels of the complex field ($i_n$) at the input FOV, which then gets modulated by a trained D²NN. For each one of the above-described schemes, $A \to A$, $I \to I$, and $P \to I$, the corresponding D²NN is trained to perform the class-specific transformations from its input FOV to the output FOV, where the input information is encoded into the corresponding channel (i.e., $A$, $I$, and $P$, respectively) of the input complex field, $i_n$ (see Table 1 and Fig. 1b). Input fields are assumed to be two-dimensional; however, in our matrix-vector multiplication notation described below, we refer to *vectorized* versions of these 2D physical quantities of phase and amplitude distributions, following the notation of earlier work[28].

For the $A \to A$ transformations, the information of the normalized input image ($G_n$) from the data class $n$ is encoded as the amplitude channel of the complex input field (i.e., $|A_n| = G_n$), whereas the phase of the input field is set to zero, i.e., $\phi_n = 0$. Therefore, the complex input optical field can be represented as $i_n = |A_n|e^{j2\pi\phi_n} = G_n$. For an input object from data class $n$, after all-optically processed by the diffractive network, the *amplitude* distribution of the resulting output optical field (i.e., $|\mathbf{D}i_n|$) approximates the linearly transformed version of the input field amplitude using the transformation matrix that is specifically assigned to that data class $n$, i.e.,

$$\alpha|\mathbf{D}i_n| \approx T_m|i_n| = T_m|A_n| = T_m G_n, \qquad \text{iff. } m = n \tag{1}$$

where $\alpha$ is a normalization scalar that takes into account the optical losses and the output diffraction efficiency of the diffractive network for each target transformation. At the decryption stage, the user can apply the key $T_m^{-1}$ to the diffractive network's output amplitude measurement, and normalize the resulting image ($T_m^{-1}|\mathbf{D}i_n|$) using its maximum value as the scaling factor, i.e. $\alpha = \frac{1}{\max\{T_m^{-1}|\mathbf{D}i_n|\}}$. Therefore, this normalization factor $\alpha$ takes care of various sources of optical losses between the input and output FOVs of the class-specific D²NN, such that we have $T_m^{-1}\alpha|\mathbf{D}i_n| \approx G_n$, enabling the decryption of the original image, $G_n$.

Similarly, for the $I \to I$ transformations, the image information is encoded as the intensity of the complex input field (i.e., $|A_n|^2 = G_n$), and the *intensity* distribution of the output optical field approximates the class-specific linear transformation of the intensity of the input optical field, i.e.,

$$\beta|\mathbf{D}i_n|^2 \approx T_m|i_n|^2 = T_m|A_n|^2 = T_m G_n, \qquad \text{iff. } m = n \tag{2}$$

where $\beta = \frac{1}{\max\{T_m^{-1}|\mathbf{D}i_n|^2\}}$ is a normalization scalar, similar to $\alpha$.

Following the same flow of logic, for the $P \to I$ transformations, the information of the image data is encoded into the phase channel of the complex input field, whereas the amplitude of the input field is set to 1, i.e., $i_n = e^{j2\pi\phi_n}$, where $\phi_n = G_n$, i.e., $\phi_n \in [0, 1]$. The *intensity* distribution of the output optical field behind the trained D²NN approximates the class-specifically transformed phase of the input complex field, i.e.,

$$\beta|\mathbf{D}i_n|^2 \approx T_m \frac{\angle i_n}{2\pi} = T_m \phi_n = T_m G_n, \qquad \text{iff. } m = n \qquad (3)$$

Figure 2 shows an application scenario of the presented class-specific all-optical transformation network. The images from $N$ different data classes are encoded into the input optical fields and then processed, one by one, by the diffractive optical network. The original images are therefore "encrypted" in a class-specific manner, resulting in non-informative measurement patterns at the output FOV for all the data classes. The original image information can be subsequently recovered by digitally applying the correct inverse linear transformation ($T_m^{-1}$, $m = n$) that corresponds to the specific data class $n$. However, when the non-matching inverse transformation ($m \neq n$) is applied, the output measurement cannot be decrypted, resulting in noise-like, meaningless features. This class-specificity feature enables sharing of the acquired image data, encrypted by a trained D²NN, with different users, where each user can only decode the images of the data class assigned to their corresponding key, $T_m^{-1}$.

We first numerically demonstrated the class-specific $A \to A$ transformations with a seven-layer diffractive network design. As shown in Fig. 3a. each diffractive layer contains 400×400 trainable diffractive features/neurons, each with a lateral size of ~0.53$\lambda$, which only modulate the phase of the transmitted wave field. The axial distance between two consecutive planes was set to 40$\lambda$. The class-specific $A \to A$ transformation network was demonstrated using the handwritten EMNIST database[24], from which four data classes, including the upper-case handwritten letters "U", "C", "L", and "A" were selected as our target data classes of interest. Each EMNIST grayscale image was first normalized to the range $[0, 1]$ and then encoded as the amplitude of the input optical field to be processed by the trained D²NN, i.e., $|A_n| = G_n$ is the normalized handwritten EMNIST image. Four different linear transformation matrices ($T_1$, $T_2$, $T_3$, and $T_4$, all real-valued, non-negative and normalized, see Supplementary Fig. 1a) were randomly generated, each of which was exclusively designated to only one of the data classes. The diffractive network for this data class-specific $A \to A$ transformation was trained using deep learning through the iterative stochastic gradient-descent optimization (see the Methods section). After its convergence, the resulting D²NN (Fig. 3c) was numerically tested using the handwritten EMNIST test letters, which were not included in the training process. In the blind testing phase summarized in Figs. 3b and 4, the resulting normalized amplitude of the diffractive network's output field closely approximates the class-specific linearly transformed versions of the input field amplitude, i.e., $\alpha|\mathbf{D}i_n| \approx T_m|i_n| = T_m|A_n|$, iff. $m = n$ using the four transformation matrices ($m = 1, 2, 3, 4$) pre-assigned to the input data classes. After being processed by the diffractive network, the original images from all four data classes were encrypted based on the class-specific transformation matrices ($T_1$, $T_2$, $T_3$, and $T_4$), as demonstrated in Figs. 3b and 4. Only by applying the corresponding inverse linear transformation ($T_m^{-1}$), the original image information can be retrieved. Taking the handwritten letter "U" as an example, the diffractive network linearly transformed and encrypted the input amplitude image using $T_1$ (i.e. $|\mathbf{D}i_1| \propto T_1|i_1|$), which can be retrieved only by applying the correct inverse transformation $T_1^{-1}$, while applying other mismatching inverse transformation matrices ($T_2^{-1}, T_3^{-1}$, and $T_4^{-1}$) resulted in uninterpretable images (see the off-diagonal images in Fig. 3b).

In Fig. 4, we further demonstrate the generalization capability of the converged $A \to A$ transformation D²NN and show more blind testing results using the target class letters in different handwriting styles. For all the different handwritten input letters, the network generated encrypted output amplitude patterns (the second column of each box in Fig. 4). By sequentially applying the inverse linear transformations ($T_m^{-1}$) to the diffractive network output amplitude, the original letters can be faithfully reconstructed only when the correct inverse transformation is selected (i.e., $m = n$, the third column of each box in Fig. 4); all the other mismatching inverse transformations still resulted in noise-like patterns (the last three columns in each box in Fig. 4). This confirms the specificity of the decryption keys ($T_m^{-1}$) to the data classes.

Next, we tested the same trained diffractive transformation network ($A \to A$) with different objects that are *not* within the four training data classes, i.e., the handwritten letters that are not "U", "C", "L", or "A" (i.e., $G_n, n \notin \{1, 2, 3, 4\}$). As shown in Supplementary Fig. 2, for all the selected handwritten test letters that are not part of our target data classes, the diffractive network generated uninterpretable outputs, and the original image information could not be retrieved no matter which inverse transformation matrix was applied to the output. These results further illustrate the success of this all-optical data class-specific $A \to A$ transformation D²NN, which learned to perform the assigned class-specific transformations *exclusively and selectively* on the objects that belong to the desired data classes even if they are in diverse object styles, while concealing the information of all the other objects outside the selected data classes and securing the original data.

Next, we trained a new diffractive network to numerically demonstrate the class-specific $P \to I$ transformation network, as shown in Fig. 5. The four handwritten EMNIST data classes, including the upper-case letters "U", "C", "L", and "A" were again selected as our target classes of interest. Each handwritten EMNIST image was first normalized to the range [0,1] and then encoded into the phase of the complex input field, whereas the amplitude of the input field was set to be 1 (i.e., $|A_n| = 1$, $\phi_n = G_n$ is the normalized handwritten EMNIST image). For each data class, the corresponding class-specific transformation matrix $T_m$ was kept the same as those used in the $A \to A$ transformation, i.e., $T_1$, $T_2$, $T_3$, and $T_4$ reported in Supplementary Fig. 1a. After its convergence, the trained $P \to I$ transformation D²NN acted as a data class-specific "lock" that encrypted the input phase information from certain data classes using pre-assigned transformation matrices, i.e., $\beta|\mathbf{D}i_n|^2 \approx T_m\phi_n$, iff. $m = n$. Only by applying the correct decryption key ($T_m^{-1}$) can the original phase images be recovered, while applying other mismatching keys or inverse transformations resulted in noise-like outcomes (see Fig. 5a). Similar to the case of $A \to A$ transformations reported earlier, this class-specific $P \to I$ transformation D²NN was further tested with handwritten letters that were *not* used in the training phase, including the EMNIST test letters within the four selected data classes ("U", "C", "L", "A") and the handwritten letters outside the four selected data classes, as shown in Fig. 6 and Supplementary Fig. 3, respectively. The network can all-optically encrypt/transform the phase-only objects in diverse handwriting styles according to their classes using the specific transformation matrices pre-assigned to each target class, outputting intensity patterns that could be decrypted only by applying the correct inverse transformations, demonstrating the generalization capability of our diffractive network. Furthermore, for the other types of objects that are not within the four selected data classes, their original phase information could not be recovered even by applying the decryption keys, as demonstrated in Supplementary Fig. 3.

Next, we applied the same framework for $I \to I$ transformations and selected the permutation operation as a test bed. As a specific form of a linear transformation, permutation is a frequently used computational operation in the fields of digital communication and data security[29–31]. We designed a class-specific permutation network implemented under the $I \to I$ transformation framework using the D²NN architecture as shown in Fig. 2a. For this task, we used ten classes from the QuickDraw[25] dataset as the data classes of interest. Correspondingly, ten arbitrarily generated permutation matrices ($P_1$, $P_2$, …, $P_{10}$, see Supplementary Fig. 1b) were individually assigned to these ten data classes as the class-specific transformations to be all-optically performed by a trained D²NN. The generation of these permutation matrices was constrained so that all the selected matrices have no overlapping non-zero entries to avoid the use of similar permutation operations; stated differently $P_i \odot P_j = \bar{0}$ for $i \neq j$, where $\odot$ refers to the Hadamard product (see the Methods section). In this numerical experiment, the QuickDraw images from the selected data classes (see Fig. 7) were normalized to the range $[0, 1]$ and encoded as the intensity of the input field, i.e., $|A_n|^2 = G_n$ is the normalized QuickDraw image to be all-optically processed by the trained D²NN. The diffractive network was trained so that the normalized output intensity matches the class-specifically permutated input image, i.e., $\beta|\mathbf{D}i_n|^2 \approx P_m|i_n|^2 = P_m|A_n|^2 = P_mG_n$, iff. $m = n$.

Similar to the earlier results reported for the $A \rightarrow A$ and $P \rightarrow I$ transformations, through this all-optical class-specific permutation operation, computed by the diffractive network, the resulting output images showed noise-like patterns that are not interpretable by human readers. Only when the correct inverse permutation was applied did the recovered images faithfully reflect the original images being encrypted, while the unpaired inverse permutation operations (for $m \neq n$) resulted in noise-like uninterpretable images, demonstrating the class-specificity of the all-optical permutation operations (covering $P_1, P_2, \ldots, P_{10}$ for 10 distinct data classes) performed by the trained $D^2NN$. Additional examples of the blind testing results are further reported in Fig. 8 and Supplementary Fig. 4; the latter reports the test objects that are not within the 10 selected data classes, where the original object information could not be recovered by applying any of the decryption permutation keys ($P_1^{-1}, P_2^{-1}, \ldots, P_{10}^{-1}$).

This framework also works for more complicated objects; to demonstrate this, we also report a $D^2NN$ design trained to all-optically perform class-specific permutation operations with $P_1, P_2, \ldots, P_{10}$ (see Supplementary Fig. 1b) individually assigned to one data class of the Fashion MNIST dataset[26]. The blind testing results of this class-specific permutation diffractive network are reported in Supplementary Fig. S5, which achieved a similar level of success as demonstrated earlier for the QuickDraw image dataset (Fig. 8). To better quantify the success of these class-specific all-optical transformations, Supplementary Fig. S6 reports the average Pearson Correlation Coefficient (PCC)[32] values between the reconstructed test images and the original input images from the QuickDraw dataset and Fashion MNIST dataset (excluded from the training phase) using the models reported in Fig. 8 and Supplementary Fig. S5, respectively. Each element of the reported confusion matrices was averaged over 100 test objects from the same data class for the QuickDraw results and 1000 test objects from the same data class for the Fashion MNIST results. These results demonstrate that the original images can be faithfully reconstructed by applying the correct class-specific inverse transformations, achieving high PCC values for $m = n$, i.e., the diagonal elements of the confusion matrices. On the other hand, when the mismatching inverse transformations were applied, the PCC values were much lower for $m \neq n$, i.e., the off-diagonal elements of the confusion matrices, which once again confirms the class-specificity and accuracy of these all-optical transformations for the $N = 10$ data classes selected from the QuickDraw (Fig. 8) and Fashion MNIST (Supplementary Fig. S5) datasets.

### 2.2. Experimental demonstration of a class-specific transformation $D^2NN$

We experimentally demonstrated the proof of concept of this class-specific all-optical image transformation scheme by building an $I \rightarrow I$ $D^2NN$ that operates at $\lambda =1550$ nm. This class-specific

D²NN was trained using the MNIST[33] dataset containing ten handwritten digits/classes. We specifically designated $T_3$ as the identity matrix ($T_3 = \bar{I}$) so that for an input image $G_n$ that belongs to the data class $n = 3$, the D²NN's output intensity pattern closely resembles the original image, i.e., $|\mathbf{D}i_3|^2 \propto T_3|i_3|^2 = \bar{I}|i_3|^2 = G_3$. For the other types of input objects $G_n$ that do not belong to the data class $n = 3$, the D²NN was designed to all-optically erase their information for $n \neq 3$. In other words, we trained a data class-specific D²NN which specifically performs the identity transformation to the handwritten digits "2" (i.e., $G_3$); see Fig. 9a.

For this experimental demonstration, we trained a two-layer diffractive network operating at $\lambda = 1550$ nm using the configuration described in Fig. 9b. To mitigate the potential misalignments during the fabrication and the experimental set-up, the D²NN was "vaccinated" during its training phase[34] by introducing random axial displacements to the diffractive layers (see the Methods section for details). The resulting trained diffractive layers are shown in Fig. 9b. After its training, the diffractive network was fabricated using two-photon polymerization-based 3D printing, and was experimentally tested using $\lambda = 1550$ nm illumination (Fig. 9c-d). Supplementary Fig. S7 shows the fabricated diffractive surfaces imaged using a scanning electron microscope (SEM). 12 different MNIST handwritten test digits that are not included in the training process were used as the input test objects, which were fabricated using photolithography to form the binary transmittance patterns. After being modulated by the diffractive network, the optical field intensity at the D²NN output FOV was captured using a charge-coupled device (CCD). Figure 9e shows the experimental testing results using the fabricated diffractive network along with their corresponding numerical simulation results. These results demonstrate that the fabricated D²NN worked as designed to perform the identity transformation to the selected input data class ($n = 3$), while all-optically erasing the information of other classes of test objects at its output FOV, further confirming the feasibility of our class-specific transformation network design.

## 3. Discussion

We reported a D²NN-based framework that performs all-optical class-specific transformations. Since the transformation/encryption process is all-optically performed on the input optical fields of the objects through light propagation across passive diffractive surfaces, it is fast, secure, and energy-efficient without requiring any external computing power. Instead of utilizing one fixed transformation matrix for all the input objects, the presented method can approximate multiple pre-assigned transformations depending on the input data classes, enabling data class-specific encryption of information. Such a class-specific design adds an additional layer of security during the data transmission, where the leakage of one encryption key would not result in the breach of all the original data. This class-specific design could also enable parallel secure data distribution to many users using only a single encryption device (D²NN). In other words, all the original information can be encrypted using one front-end encryption network and transferred to users at the same time. Depending on the data access permissions assigned to individual users, different decryption keys (inverse transformation matrices) can be distributed to the different receivers so that only the desired portion of the data is shared with the authorized end receivers/users. Moreover, even though the decryption process in our work is achieved by digital approaches, it could also be designed as another optical computing device (another D²NN) by training a symmetric inverse-transformation network. In this case, the encryptor and decryptor D²NNs could be assembled to, e.g., two ends of the transmission link, to achieve all-optical class-specific secure data transmission, i.e., all-optical D²NN-based encoding and all-optical D²NN-based decoding, one following another.

Although a coherent diffractive optical processor is linear in complex fields, the trained class-specific diffractive network *approximates a nonlinear operation* applied to the input complex-valued optical fields. For instance, in the $P \to I$ transformation scheme, the intensity of the output field ($|\mathbf{D}i_n|^2$) is proportional to the linearly transformed version of the phase of its input field, i.e.,

$$|\mathbf{D}i_n|^2 \propto T_m \phi_n, \qquad \text{iff. } m = n \tag{4}$$

$$|\mathbf{D}i_n| \propto \sqrt{T_m \phi_n} = \sqrt{\frac{T_m \angle i_n}{2\pi}}, \qquad \text{iff. } m = n \tag{5}$$

which is in fact a nonlinear operation that is applied to the optical input fields ($i_n$), and this nonlinear operation is approximated by the trained D²NN.

Similarly, $A \to A$ or $I \to I$ based transformations defined in our work are also approximating nonlinear operations applied on the optical input fields, $i_n$. This is easy to see since in general $|D^2NN\{|a| + |b|\}| \neq |D^2NN\{|a|\}| + |D^2NN\{|b|\}|$ or $|D^2NN\{|a| + |b|\}|^2 \neq |D^2NN\{|a|\}|^2 + |D^2NN\{|b|\}|^2$. A diffractive optical processor that operates with spatially and temporally coherent fields is linear in complex fields, and therefore $D^2NN\{|a| + |b|\} = D^2NN\{|a|\} + D^2NN\{|b|\}$ or in general, $D^2NN\{a + b\} = D^2NN\{a\} + D^2NN\{b\}$.

## 4. Methods

### 4.1. Generation of linear transformation matrices

The 2D linear transformation matrices $T_1$, $T_2$, $T_3$, and $T_4$ (reported in Supplementary Fig. 1a) have the size of $N_i \times N_o = 225 \times 225$, each representing a unique mapping from the input images with $N_i = 15 \times 15$ pixels to the output images with $N_o = 15 \times 15$ pixels. Each transformation matrix was generated by assigning 100 random entries of each row with random values sampled from the uniform distribution $\mathbf{U}[0, 1]$, and the remaining entries were assigned to zero values. Then the transformation matrix was normalized so that the summation of each row is 1. To avoid ill-conditioned transformation matrices and ensure accurate computation of their inverse (for the decryption phase), we only kept the matrices with condition numbers less than 1000. After generating the linear transformation matrices, we also evaluated the cosine similarity values between any two matrices (see Supplementary Fig. 1c):

$$\text{cosine similarity}(T_i, T_j) = \frac{vec(T_i) \cdot vec(T_j)}{\|vec(T_i)\| \, \|vec(T_j)\|} \tag{6}$$

where $vec(T_i)$ and $vec(T_j)$ are the vectorized counterparts of the matrices $T_i$ and $T_j$, respectively. The cosine similarity values between any two transformation matrices were calculated to be around 0.33, showing the low degree of likeness of the randomly generated transformation matrices.

The 2D permutation matrices $P_1$, $P_2$, …, $P_{10}$ (reported in Supplementary Fig. 1b) have the size of $N_i \times N_o = 784 \times 784$, each representing a pixel-wise permutation mapping from the input images with $N_i = 28 \times 28$ pixels to the output images with $N_o = 28 \times 28$ pixels. Each of the permutation matrices was generated by randomly shuffling the rows of a 784×784 identity matrix. We also enforced that all the permutation matrices have no overlapping non-zero entries, i.e., $P_i \odot P_j = \bar{0}$ for $i \neq j$, in order to avoid similar permutation operations being used. Therefore, the cosine similarity values between any of the two permutation matrices used in this work are all zero (see Supplementary Fig. 1c).

### 4.2. Optical forward model of the class-specific transformation D²NN

For a diffractive network consisting of multiple successive diffractive layers, every two consecutive layers are connected to each other by free-space propagation of the light field. In our work, the free-space propagation is modeled through the angular spectrum approach[4], in which an optical field $u(x, y, z)$ located at $z = z_a$ being propagated for a distance of $d$ along the axial direction (i.e., along the z axis) can be written as[35]:

$$u(x, y, z = z_a + d) = \mathcal{F}^{-1}\{\mathcal{F}\{u(x, y, z = z_a)\} H(f_x, f_y; d)\} \tag{7}$$

where $\mathcal{F}$ is the 2D Fourier transform and $\mathcal{F}^{-1}$ is the 2D inverse Fourier transform operation. $f_x$ and $f_y$ are the spatial frequencies along the $x$ and $y$ (lateral) directions, respectively. $H(f_x, f_y; d)$ is the transfer function of free space, defined as:

$$H(f_x, f_y; d) = \begin{cases} \exp\left\{jkd\sqrt{1 - \left(\frac{2\pi f_x}{k}\right)^2 - \left(\frac{2\pi f_y}{k}\right)^2}\right\}, & f_x^2 + f_y^2 < \frac{1}{\lambda^2} \\ 0, & f_x^2 + f_y^2 \geq \frac{1}{\lambda^2} \end{cases} \quad (8)$$

where $k = \frac{2\pi}{\lambda}$ is the angular wavenumber of the illumination light ($\lambda$).

For the light field that reaches each diffractive layer, we consider only the phase modulation of the transmitted field. We selected $\Phi(x, y, z_l)$ as the trainable physical parameter, which represents the phase modulation value of the diffractive neuron located on the $l^{th}$ diffractive layer at $(x, y, z_l)$. The transmittance coefficient $t(x, y, z_l)$ of the corresponding diffractive neuron can be written as:

$$t(x, y, z_l) = \exp\{j\Phi(x, y, z_l)\} \quad (9)$$

Given any input optical field $i(x, y, z = 0)$, the corresponding complex optical field at the output plane $u_{out}(x, y, z = z_{out}) = \mathbf{D}i(x, y, z = 0)$ can be obtained following a sequence of free-space propagation (equation (7)) and modulation of diffractive layers (equation (9)). Given an image $G(x, y)$ (normalized to the range $[0, 1]$) to be transformed, the input optical field, $i = |A|e^{j2\pi\phi}$, can be expressed based on different encoding schemes:

Amplitude encoding:

$$i(x, y, z = 0) = |A(x, y)| = G(x, y) \quad (10)$$

Intensity encoding:

$$i(x, y, z = 0) = |A(x, y)| = \sqrt{G(x, y)} \quad (11)$$

Phase encoding:

$$i(x, y, z = 0) = \exp\{j2\pi\phi(x, y)\} = \exp\{j2\pi G(x, y)\} \quad (12)$$

After the diffractive network's optical modulation, the measurement ($M$) of the output field can be written as:

Measurement of the amplitude:

$$M = |u_{out}(x, y, z = z_{out})| = |\mathbf{D}i(x, y, z = 0)| \quad (13)$$

Measurement of the intensity:

$$M = |u_{out}(x, y, z = z_{out})|^2 = |\mathbf{D}i(x, y, z = 0)|^2 \tag{14}$$

**Supplementary Information file contains additional methods sections on:**

- Training loss function
- Diffractive network design, training, and evaluation
- Experimental demonstration of a class-specific D²NN

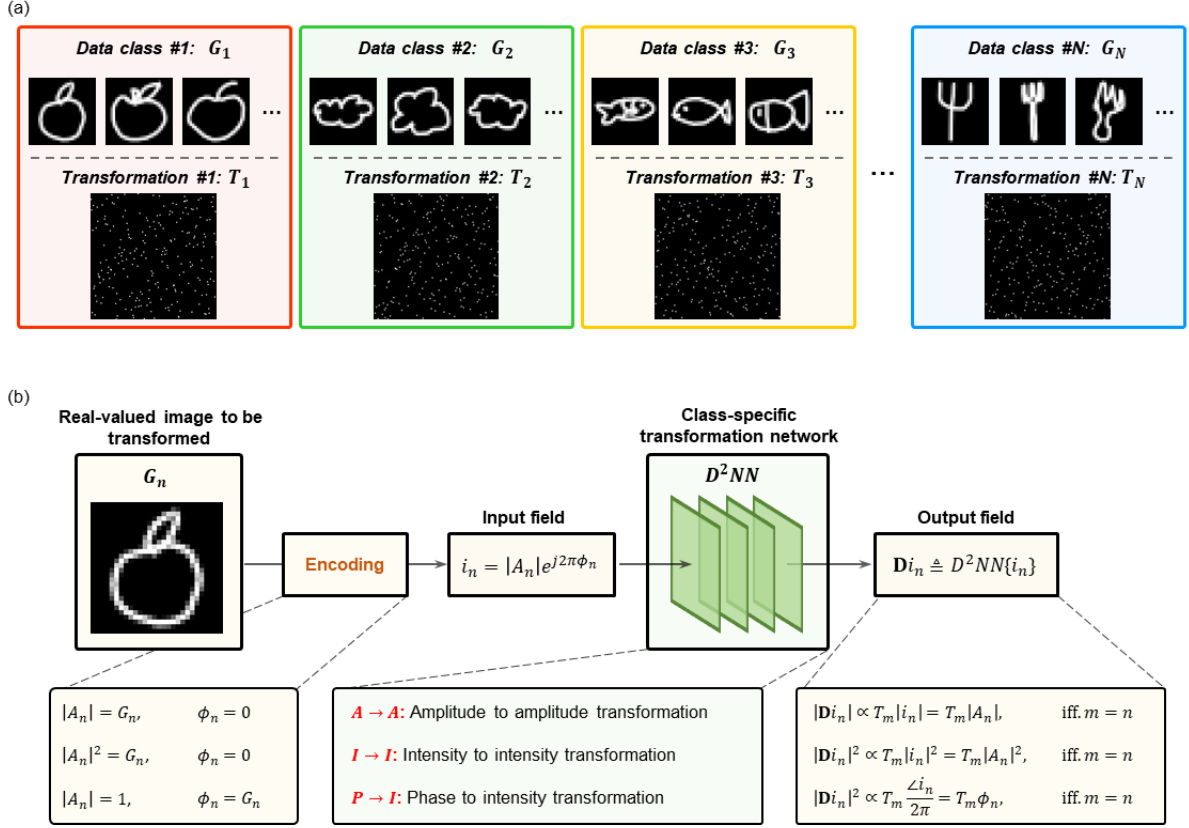

**Figure 1. Data class-specific transformations using a diffractive deep neural network (D²NN).** (a) Data from different classes are assigned to different transformation matrices, where in general $T_1 \neq T_2 \neq T_3 \neq \cdots \neq T_N$. (b) Illustration of the class-specific transformation network workflow and the different encoding/decoding schemes. Each transformation matrix, $T_m$ $(m = 1, 2, \ldots, N)$ acts on the encoded information within $G_n$ (a real-valued normalized image to be transformed/encrypted) if and only if (iff) $m = n$. Stated differently, different data classes go through different transformations using the same D²NN design.

| **A → A: Amplitude-to-amplitude transformation** | | |
|---|---|---|
| Input | $\|A_n\| = G_n,\quad \phi_n = 0,\quad i_n = \|A_n\| = G_n$ | |
| Measurement | $\alpha\|\mathbf{D}i_n\| \cong T_m\|i_n\| = T_m\|A_n\| = T_m G_n$ | iff. $m = n$ |
| Decryption | $T_m^{-1}\|\mathbf{D}i_n\| \cong T_m^{-1}T_m G_n = G_n$ | iff. $m = n$ |

| **I → I: Intensity-to-intensity transformation** | | |
|---|---|---|
| Input | $\|A_n\|^2 = G_n,\quad \phi_n = 0,\quad i_n = \|A_n\| = \sqrt{G_n}$ | |
| Measurement | $\beta\|\mathbf{D}i_n\|^2 \cong T_m\|i_n\|^2 = T_m\|A_n\|^2 = T_m G_n$ | iff. $m = n$ |
| Decryption | $T_m^{-1}\|\mathbf{D}i_n\|^2 \cong T_m^{-1}T_m G_n = G_n$ | iff. $m = n$ |

| **P → I: Phase-to-intensity transformation** | | |
|---|---|---|
| Input | $\|A_n\| = 1,\quad \phi_n = G_n,\quad i_n = e^{j2\pi\phi_n} = e^{j2\pi G_n}$ | |
| Measurement | $\beta\|\mathbf{D}i_n\|^2 \cong T_m \dfrac{\angle i_n}{2\pi} = T_m \phi_n = T_m G_n$ | iff. $m = n$ |
| Decryption | $T_m^{-1}\|\mathbf{D}i_n\|^2 \cong T_m^{-1}T_m G_n = G_n$ | iff. $m = n$ |

Normalization factor: $\alpha = \dfrac{1}{\max\{T_m^{-1}\|\mathbf{D}i_n\|\}}\qquad \beta = \dfrac{1}{\max\{T_m^{-1}\|\mathbf{D}i_n\|^2\}}$

**Table 1.** Different types of data class-specific transformations using a D²NN between an input and output FOV. These were selected since the resulting output transformations can be recorded by a simple image sensor-array.

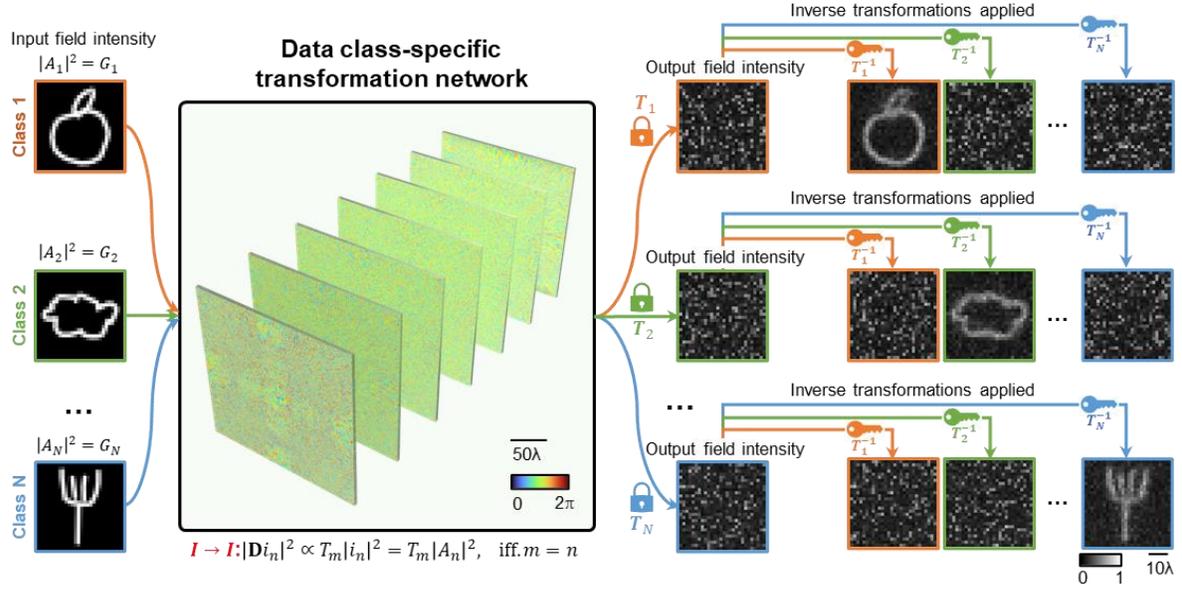

**Figure 2. Class-specific $I \to I$ transformations using a D²NN**. The intensity image information is encrypted by the diffractive neural network, in a class-specific manner, which can be retrieved at the output by applying the correct inverse transformation, key, i.e., $T_m^{-1}$ ($m = 1, 2, ..., N$). There are $N$ different transformation matrices, one assigned to each data class. As a result, $N$ different users can each be given one of the keys ($T_m^{-1}$) to one of the data classes, letting them only decrypt the images of one class of objects, while being unable to decode the other classes of objects that are transmitted.

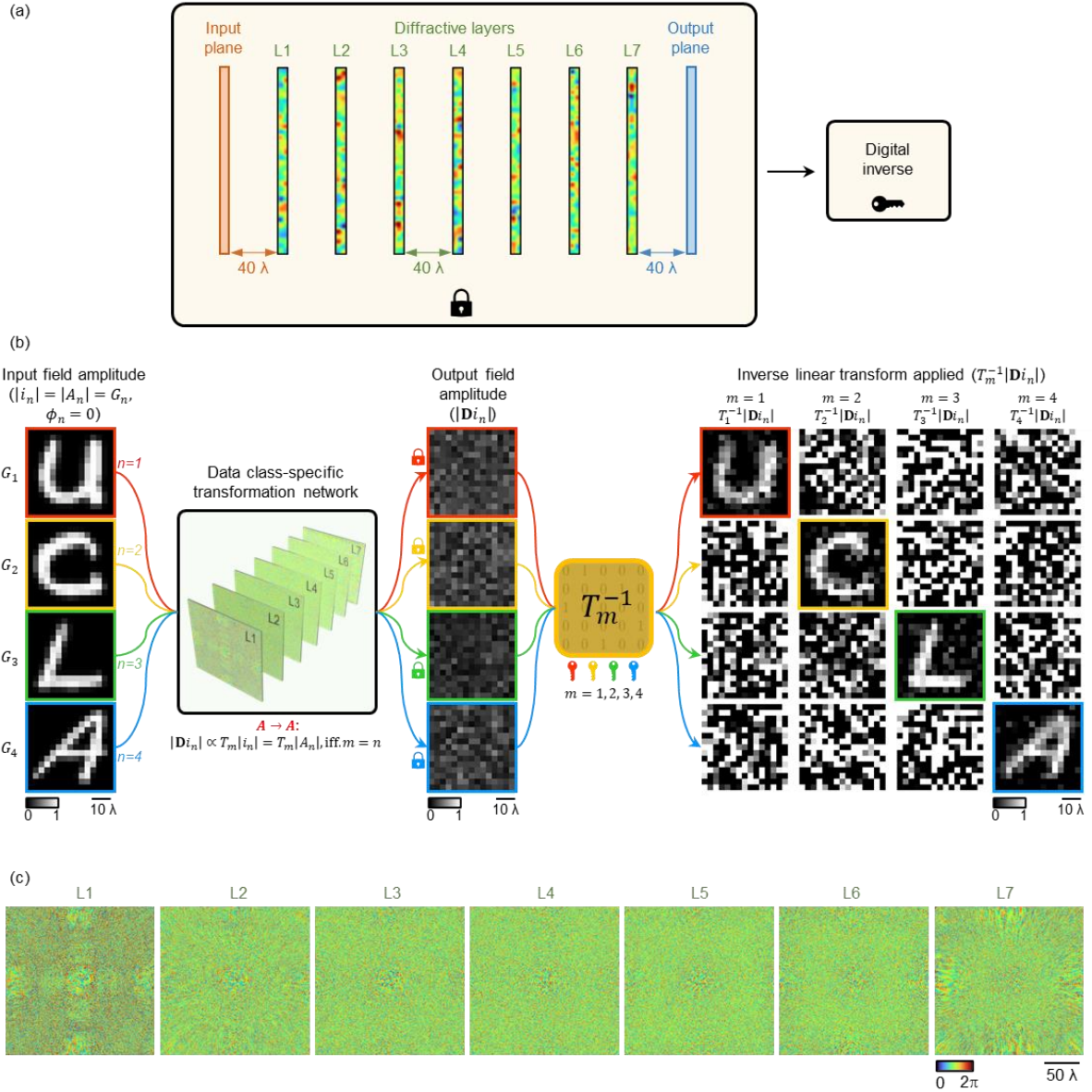

**Figure 3. Demonstration of a class-specific $A \rightarrow A$ transformation using a D²NN design.** (a) The physical layout of the seven-layer diffractive optical network. (b) The blind testing results of the trained class-specific transformation D²NN using objects encoded as the input field amplitude. There are 4 different random transformation matrices ($T_1, \ldots, T_4$), assigned to handwritten letters U, C, L, and A, respectively (see Supplementary Figure 1 for these matrices). The original image information is retrieved only by applying the matching inverse transformation that is also class-specific, $T_m^{-1}$. The reconstructed images were scaled using the same constant for visualization. (c) Trained phase modulation patterns of the diffractive layers.

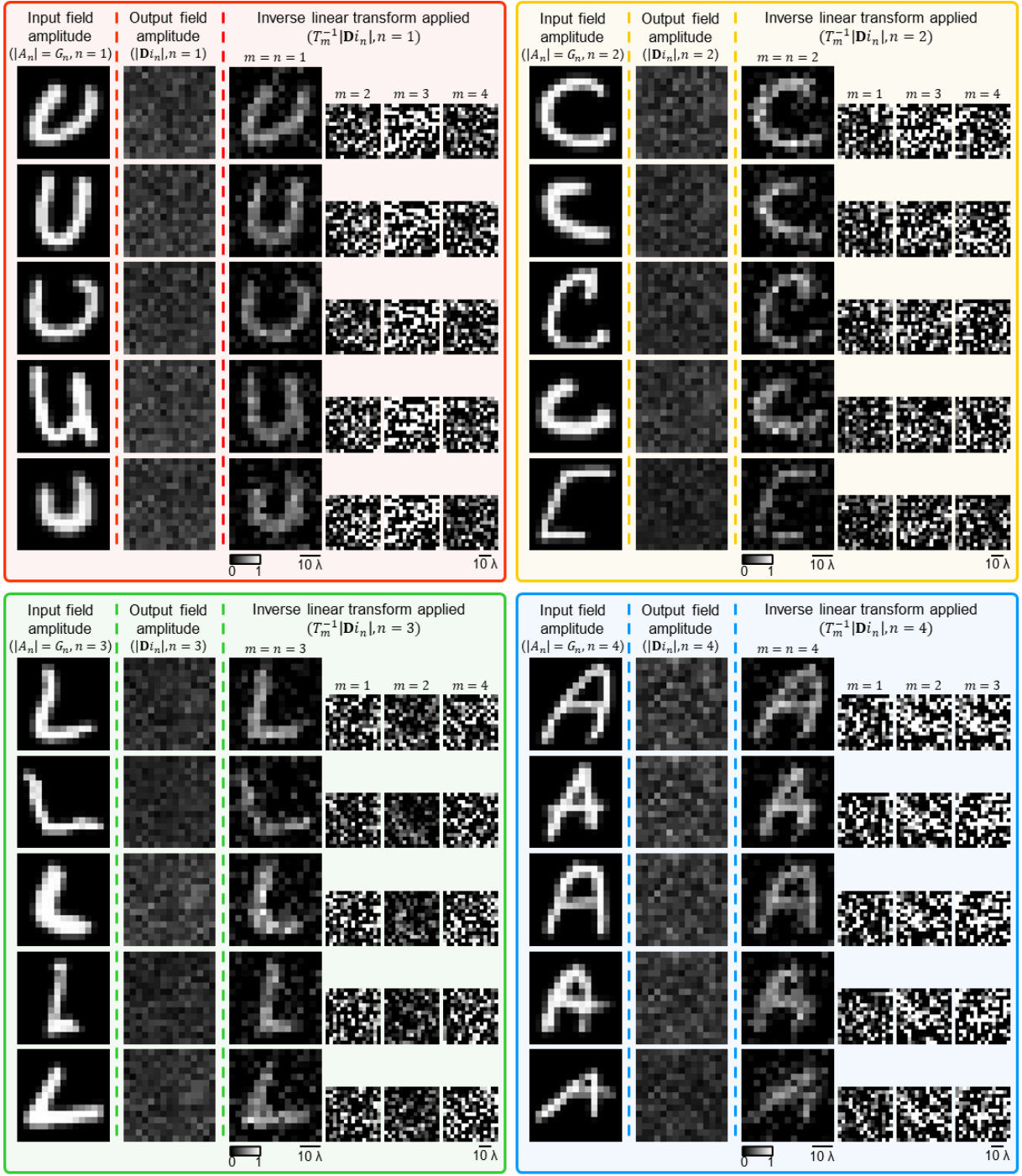

**Figure 4. Additional examples of blind testing results of the class-specific $A \to A$ transformation $D^2NN$ design shown in Fig. 3.** The reconstructed images were scaled using the same constant for visualization.

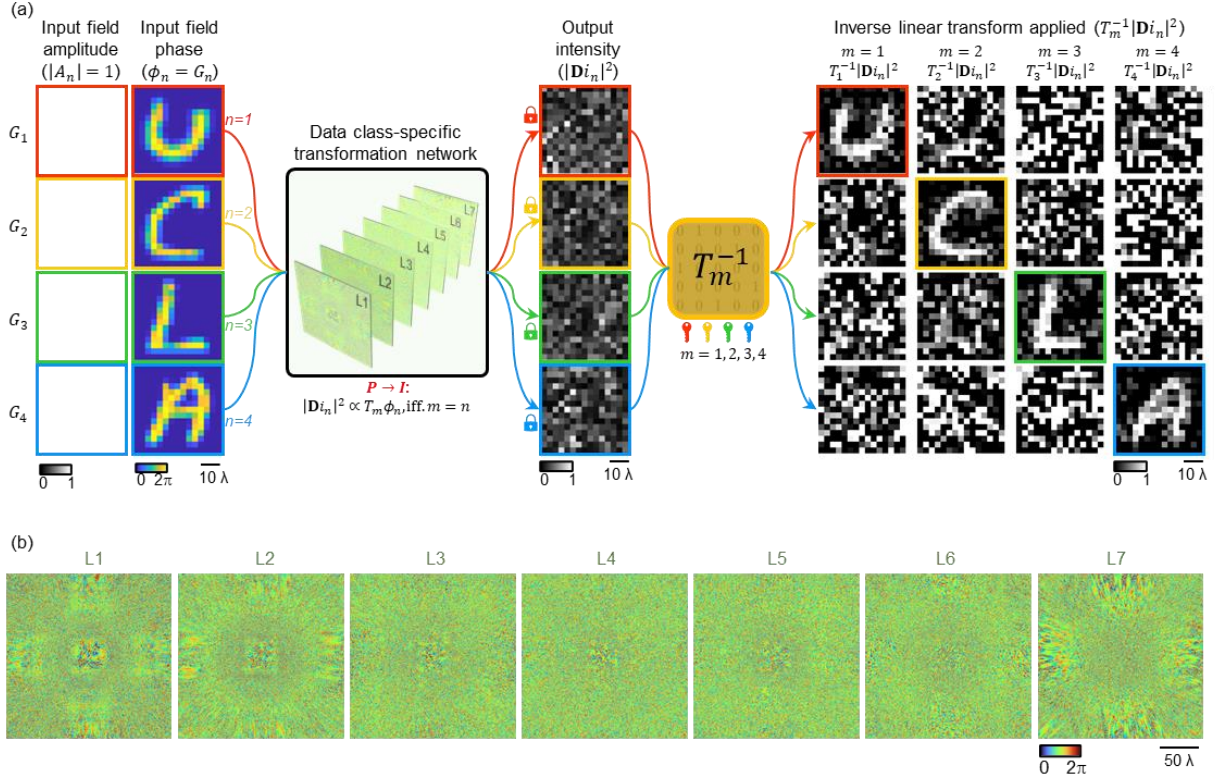

**Figure 5. Demonstration of a class-specific $P \rightarrow I$ transformation using a D²NN design.** (a) The blind testing results of the trained class-specific transformation diffractive network using objects encoded as the input field phase. There are 4 different random transformation matrices ($T_1, \ldots, T_4$), assigned to handwritten letters U, C, L, and A, respectively (see Supplementary Figure 1 for these matrices). The original phase image information is retrieved only by applying the matched inverse transformation that is also class-specific, $T_m^{-1}$. The reconstructed images were scaled using the same constant for visualization. (b) Trained phase modulation patterns of the diffractive layers.

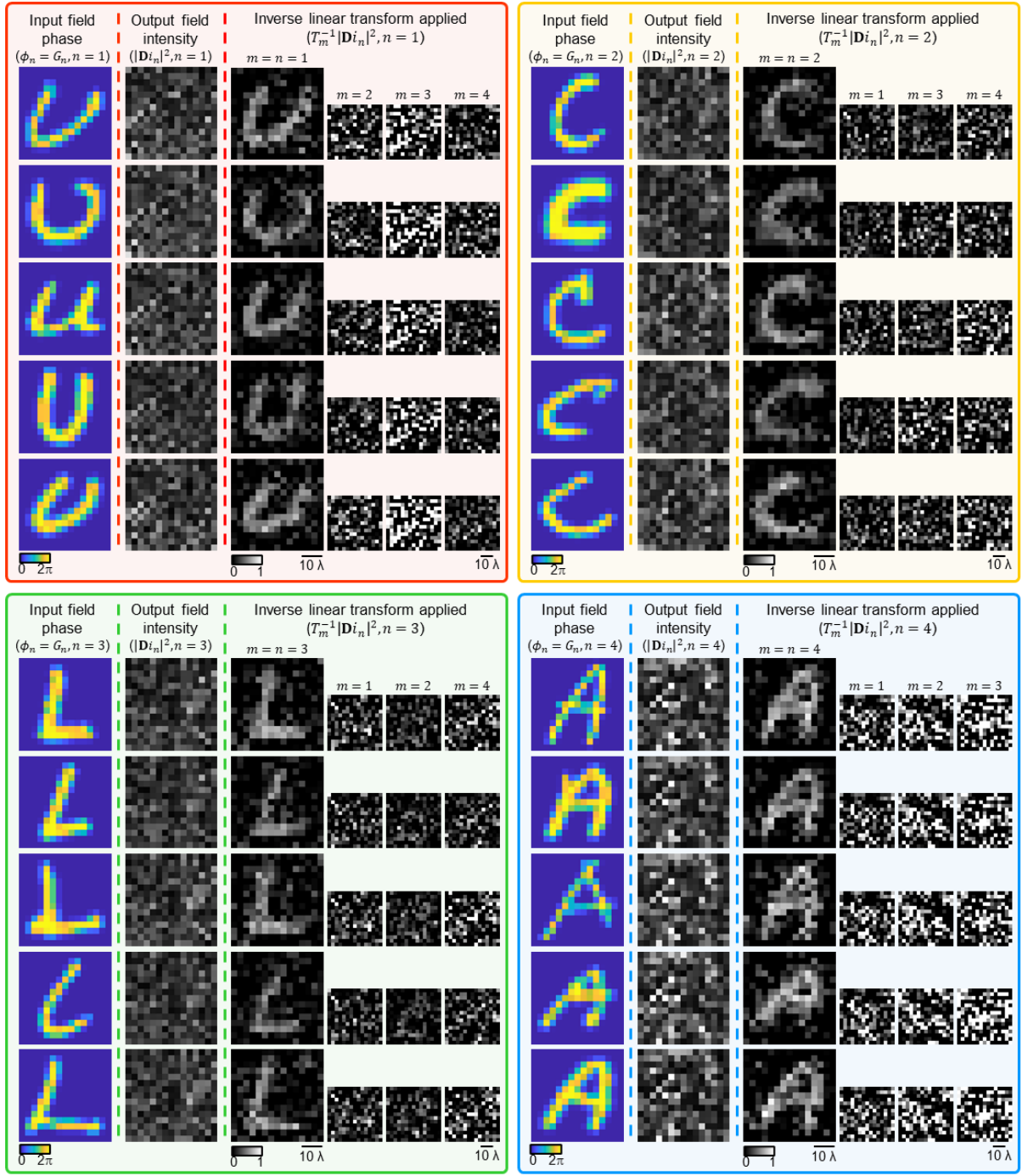

**Figure 6. Additional examples of blind testing results of the class-specific $P \rightarrow I$ transformation $D^2NN$ design shown in Fig. 5.** The reconstructed images were scaled using the same constant for visualization.

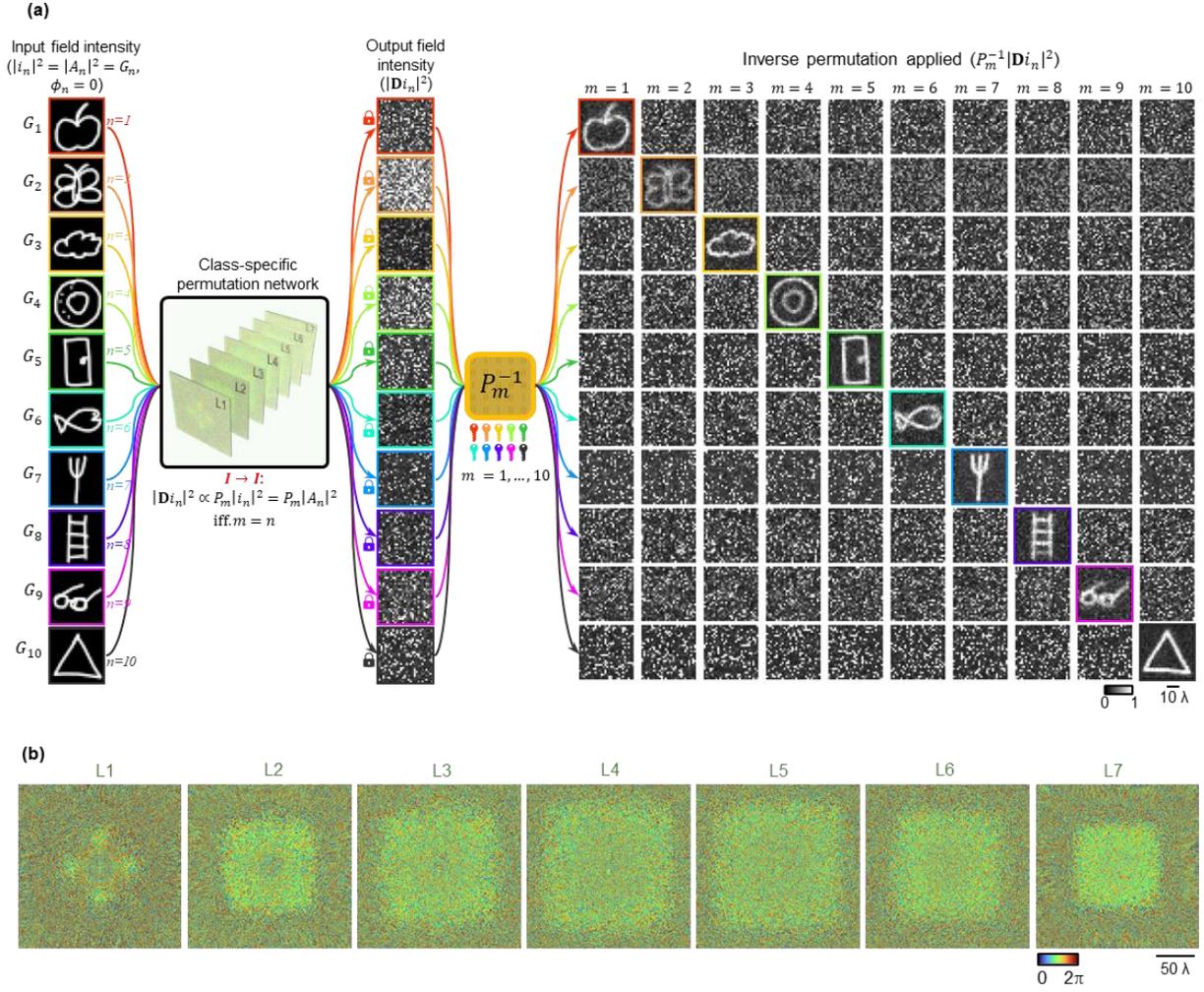

**Figure 7. Demonstration of a class-specific $I \rightarrow I$ permutation D²NN design.** (a) The blind testing results of the trained class-specific permutation network using objects encoded as the input field intensity. There are 10 different random permutation matrices ($P_1, \ldots, P_{10}$), assigned to 10 different data classes selected from the QuickDraw dataset[25]. The original image information is retrieved only by applying the matching inverse transformation that is also class-specific, $P_m^{-1}$. The reconstructed images were scaled using the same constant for visualization. (b) Trained phase modulation patterns of the diffractive layers. See Supplementary Figure 1 for these *N*=10 permutation matrices.

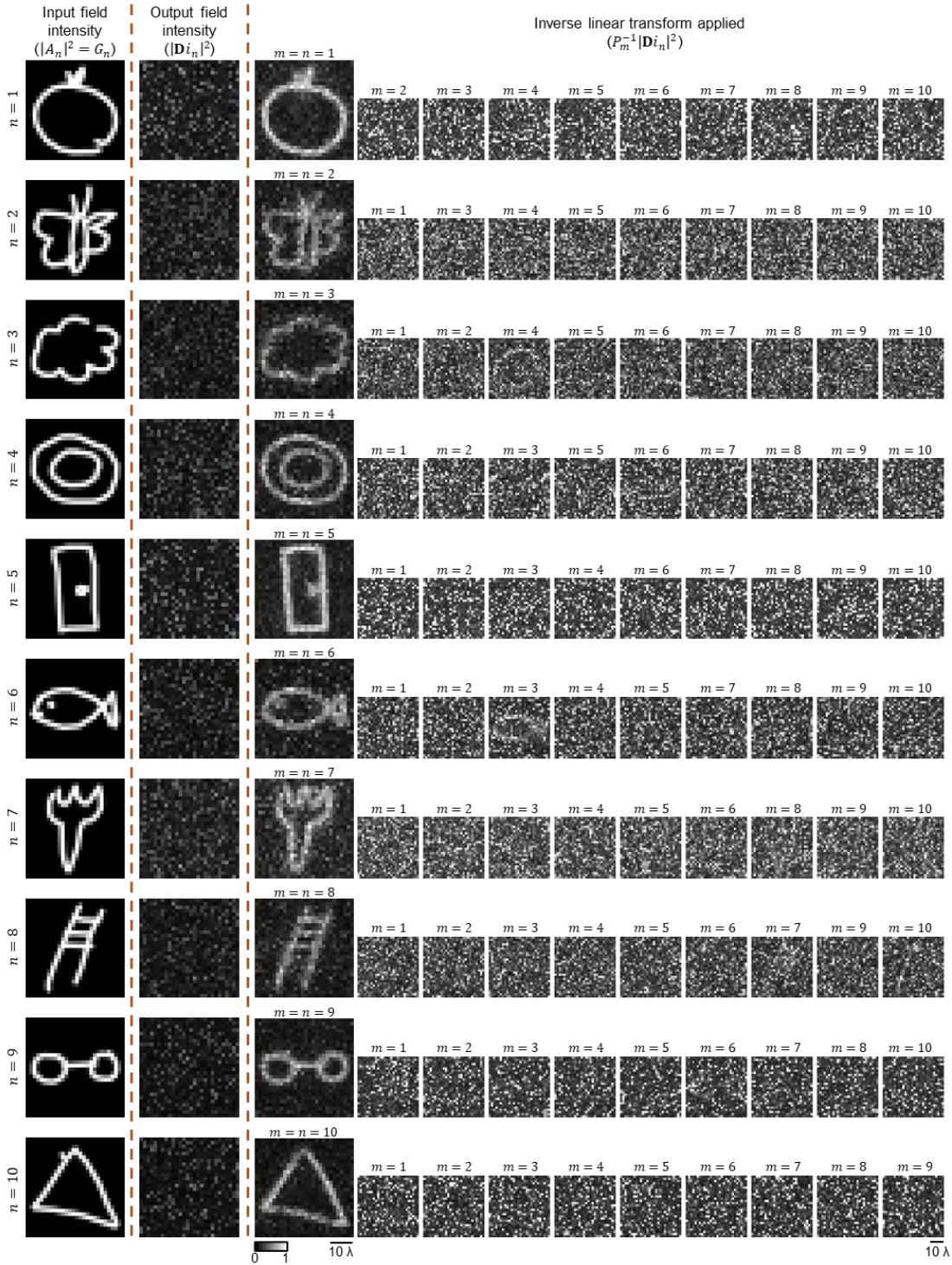

**Figure 8. Additional examples of blind testing results of the class-specific $I \rightarrow I$ permutation D²NN design shown in Fig. 7.** The reconstructed images were scaled using the same constant for visualization.

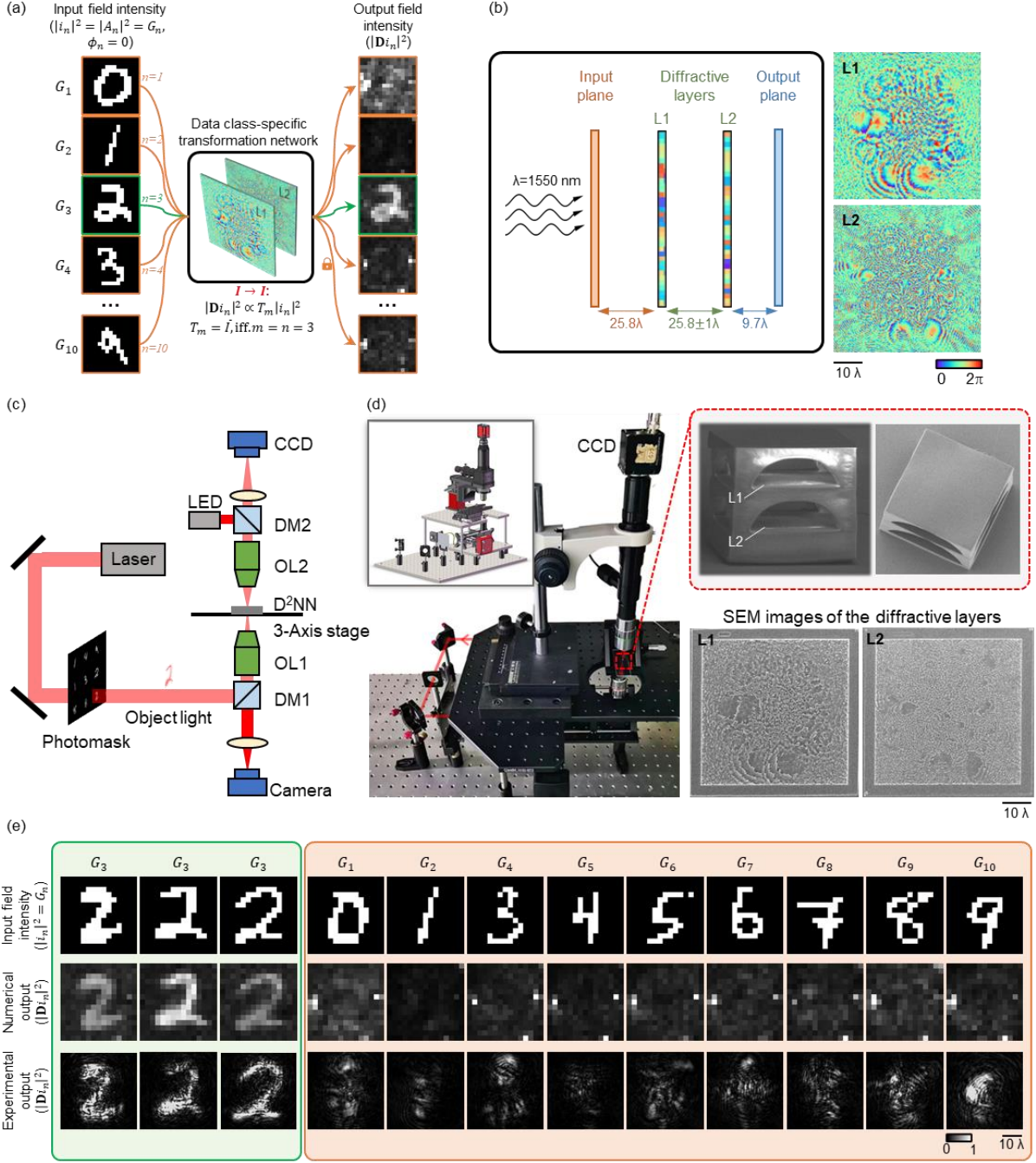

**Figure 9. Experimental demonstration of the data class-specific $I \to I$ transformation D$^2$NN.** (a) Illustration of a two-layer D$^2$NN trained to perform identity transformation to the input data class $n = 3$, while all-optically erasing other classes ($n \neq 3$) of objects at its output FOV. (b) The physical layout of the two-layer diffractive network trained under $\lambda$ =1550 nm illumination. (c) Schematic of the experimental setup using $\lambda$ =1550 nm illumination. (d) Photographs of the experimental setup and the fabricated diffractive network. (e) Experimental results of the data class-specific transformation D$^2$NN. The images in each row were scaled using the same constant for visualization.

27